\documentclass[conference]{IEEEtran}
\usepackage{amssymb,amsmath,amsthm}
\usepackage{graphicx}
\usepackage{caption}
\usepackage{subcaption}
\usepackage{mdwtab}
\usepackage{soul}
\usepackage{cite}
 \usepackage{multirow}
 \usepackage{color} 
  \usepackage[linesnumbered,ruled]{algorithm2e}
\usepackage{setspace}
\usepackage{esvect}
\usepackage{amsmath}
\usepackage{booktabs}
\usepackage{tabularx}
\usepackage{adjustbox}
\usepackage{helvet}

\usepackage{tikz}
\newcommand*\circled[1]{\tikz[baseline=(char.base)]{
            \node[shape=circle,draw,inner sep=0.3pt] (char) {#1};}}

\begin{document}
%
\title{
A Reliability Study of Parallelized VNF Chaining
}
\author{\IEEEauthorblockN{Anna Engelmann and Admela Jukan
}
\IEEEauthorblockA{Technische Universit\"at Carolo-Wilhelmina zu
Braunschweig, Germany\\
Email: \{a.engelmann, a.jukan
\} @tu-bs.de}
}
\maketitle

\begin{abstract}
In this paper, we study end-to-end service reliability in Data Center Networks (DCN) with flow and Service Function Chains (SFCs) parallelism. In our approach, we consider large flows to i) be split into multiple parallel smaller sub-flows; ii) SFC along with their VNFs are replicated into at least as many VNF instances as there are sub-flows, resulting in parallel sub-SFCs; and iii) all sub-flows are distributed over multiple shortest paths and processed in parallel by parallel sub-SFCs. We study service reliability as a function of flow and SFC parallelism and placement of parallel active and backup sub-SFCs within DCN. Based on the probability theory and by considering both server and VNF failures, we analytically derive for each studied VNF placement method the probability that all sub-flows can be successfully processed by the parallelized SFC without service interruption. We evaluate the amount of backup VNFs  required to protect the parallelized SFC with a certain level of service reliability. The results show that the proposed flow and SFC parallelism in DCN can significantly increase end-to-end service reliability, while reducing the amount of backup VNFs required, as compared to traditional SFCs with serial traffic flows. 
\end{abstract}

\IEEEpeerreviewmaketitle

\section{Introduction}
The deployment of Network Function Virtualization (NFV) in Data Center Networks (DCN) has given rise to new approaches to network load balancing, considering especially the related routing protocols. When deploying Equal Cost Multipath Protocol (ECMP), for instance, the smaller size flows ("mice") may end up being queued behind the large flows ("elephant"), while at the same large flows may compete for the same link due to hash collision, resulting in either congested or underutilized links in DCN.  Such load imbalances have been intensely studied, and among others, it has been already shown that breaking large flows into parallel sub-flows and uniformly distributing the resulting sub-flows over the network can improve the throughput, server utilization, load balancing and reducing flow completion time (FCT) without causing out-of-order problem (OOP) \cite{Xu:2014, Chakraborty:2016}. Redistribution of large and small flows over multiple paths with VNFs replicas optimally placed in DCN has been also used to improving load balancing and server utilization in DCNs \cite{Mallik:2014, Carpio:2016}.

In this paper, we ask the question of the resulting end-to-end service reliability in a system that deploys flow parallelism and VNF replications. This question is widely open as the current standards and solutions primarily consider the SFC provisioning with serial traffic flows \cite{REL-ETSI, Herker:2015,Qu:2016,Ding:2017, Hmaity:2016, Ye:2016}. Moreover, most methods proposed so far consider reliability of the individual system components, such as server, VM, link or switch, and optimize service reliability by solving reliable SFC mapping problem. This is because the consideration of multiple reliability factors is likely complex, and thus impractical.  To protect active VNFs within parallelized SFCs, additional backup VNFs, i.e., additional replicas, can replace a failed active VNF of the same type. However, this requires  server and network over-provision. On the other hand, VNF migration methods can also be used but require time to activate, and would thus lead to delays due to service interruption and network reconfigurations as well as traffic loss \cite{REL-ETSI}. 
 

In this paper, we analytically study the end-to-end service reliability in DCNs that deploys flow and VNF parallelism, - which we jointly refer to as \emph{parallelized VNF chaining}, with consideration of VM and server failures. Based on the probability theory and by assuming that the reliability of the individual VMs and servers are known, we derive end-to-end service reliability for various placement strategies of all active and backup VNFs in an SFC. We consider placement strategies of the backup and active SFCs as concentrated in one server, or distributed over the network. The  expressions derived allow us to evaluate an improvement in service reliability achieved by the backup SFC, compared to the reliability achieved without a backup. We analyze the amount of backup VNFs required for protection of parallelized SFCs to reach the required level of reliability. The results show that independently of failure protection method, the flow and VNF parallelism in DCN can significantly increase reliability and reduce amount of backup VNFs required as compared to amount of backup resources used when serial traffic flow is used for the same SFC. To the best of our knowledge, this is the first study to analyze the end-to-end service reliability in DCNs with SFC parallelism.


\par The rest of the paper is organized as follows. Section 2 presents the reference architecture based on CORD. Section 3 presents the reliability analysis. Numerical results are shown in Section 4. Section 5 concludes the paper.

\section{Parallelized SFC and Backup Strategies}
\begin{figure}[!t]
\centering
\includegraphics[width=1\columnwidth]{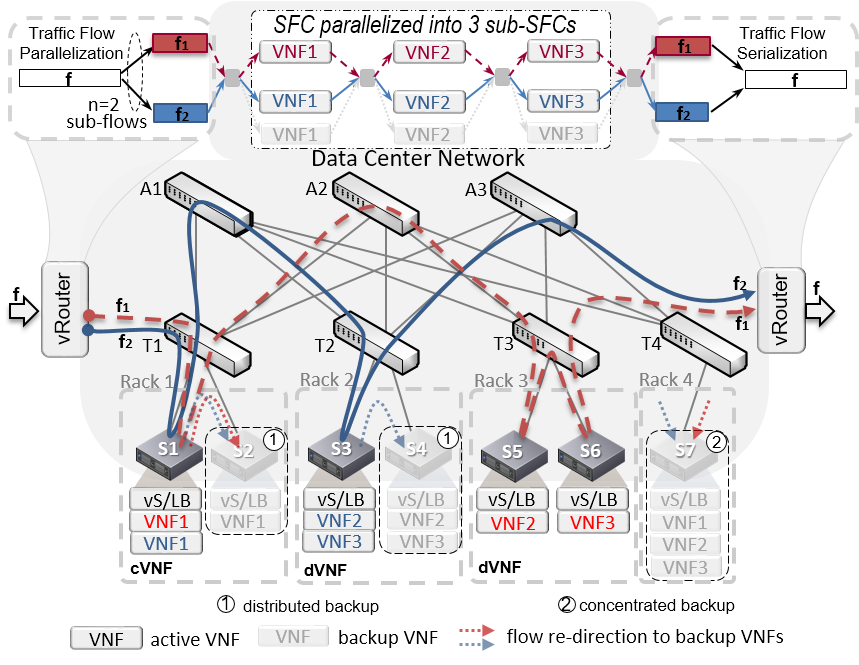}%
  \vspace{-0.1cm}
  \caption{CORD network \cite{CORD} with parallelized flow and SFC.}
  \label{net}
\end{figure}

Fig \ref{net} presents the reference CORD network, where the underlay hardware fabric consists of racks with multiple servers, switches and links. ToR switches handle forwarding within a rack, e.g., T1 handles traffic within Rack1. Forwarding across racks is performed by the so-called spine switches (A1-A3).  Each server can host multiple VMs and virtual switch (vS), e.g., programmable hypervisor switch. For a parallelized SFC, we assume that each VM reserves as many resources as required for one VNF to serve a sub-flow, i.e, each VM allocates only one VNF and   failure of any VM causes  failure of one VNF. All VMs are connected to vS, which provides load balancer (LB), whereby the main task of vS/LB is to select a VM instance, i.e., VNF, and to provide a connection to ToR switch.  We assume that DCN deploys flow and SFC parallelism, whereby i) the serial traffic flow (elephant flow) \textbf{$f$} is split into $n$ parallel sub-flows (mice flows), i.e., $f_1$ and $f_2$; ii) all VNFs of a certain SFC are replicated into as many active VNF instances as there are sub-flows, resulting in \emph{parallelized} SFC, consisting of parallel sub-SFCs; and iii) all $n$ sub-flows are independently transmitted and processed in parallel by $n$ parallel sub-SFCs. We distinguish between two possible placement strategies of each group of the same active VNF replicas: cVNF and dVNF. cVNF concentrates all replicas of a certain VNF within the same server, such as VNFs1 in server S1. dVNF in contrary distributes all replicas of the same type over different servers, see VNFs2 and VNFs3 distributed over S3, S5 and S3, S6, respectively. As a result,  sub-flow $f_1$ is processed by active VNFs in servers S1, S5 and S6 and sub-flow $f_2$ is processed by active VNFs in servers S1 and S3, resulting in a parallel traffic distribution over DCN and servers. 

Without any additional resources, a parallelized SFC is characterized by a certain level of reliability, which depends on the reliability of the underlying VM and servers. To additionally protect the active VNFs over all parallel active sub-SFCs, and thus improve reliability, a parallelized SFC can be enhanced by parallel backup sub-SFCs. The backup VNFs can replace any failed active VNF of the same type over any sub-flow. Just like active VNFs, also a backup sub-SFC can be either distributed over multiple servers \circled{\small1}, e.g, S2 and S4, or concentrated in one backup server \circled{\small2}, i.e., S7. In this paper, we consider possible failures of servers and VMs, i.e., VNFs, whereby switches and network links are assumed as highly reliable. We assume that per default active and backup VNFs are never hosted in the same server. 


As shown in \cite{REL-ETSI, Herker:2015}, to provide highly reliable communication, the most effective VNF deployment is to utilize multiple switches, e.g., vS or ToR switch, that can detect failure in hard- and software and redirect traffic to the corresponding available backup VNF, if an active server or VNF  fails as shown in Fig. \ref{net}, where SFC is parallelized into $n=2$ active and one backup sub-SFCs. The traffic toward another rack is routed by the source ToR, which addresses the ToR in the destination rack. Then traffic flows are sent to spine switches, which perform only header lookups to route the traffic to the destination ToR switch. We assume that the source ToR switch is able to change an end-to-end path utilized by sub-flows over the network, which is an important capability of CORD network utilized in this study \cite{CORD}. It should be noted that when flow and SFC parallelism are deployed, the synchronization between VNFs of the same type is necessary. To this end, an external state repository can store internal states of VNFs. In this case, any backup VNF would start its operation from the "reset" state and, then, retrieve any critical state stored by the failed original VNF from the external state repository, and that up to the point of its failure as discussed in \cite{REL-ETSI}.


\subsection{Backup Placement Strategies}
In this paper, we propose to study four different backup placement strategies, as a result of combination of the placement strategies from Fig. \ref{net} for active VNFs, i.e, cVNF and dVNF, and backup VNFs, i.e., \circled{\small1} and \circled{\small2}. We refer to these strategies as: (i) server-network (aSbN) ii) server-server (aSbS) iii)  network-network (aNbN) and iv) network - server (aNbS), whereby $a$ refers to \emph{active}, $b$ refers to \emph{backup}, and S refers to strategy where VNFs are concentrated in a server, while N indicates that VNFs are distributed over the DCN. 


\begin{figure}[!t]
\centering
\begin{subfigure}[b]{0.9\columnwidth}
\includegraphics[width=0.9\columnwidth]{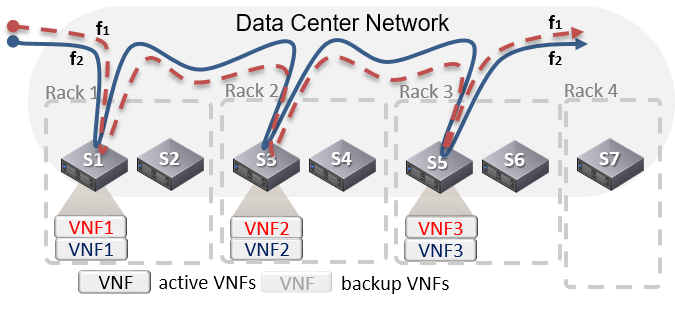}%
  \vspace{-0.2cm}
  \caption{cVNF without backup.}
  \vspace{-0.2cm}
  \label{netC0}
\end{subfigure}
\begin{subfigure}[b]{0.9\columnwidth}
\includegraphics[width=0.9\columnwidth]{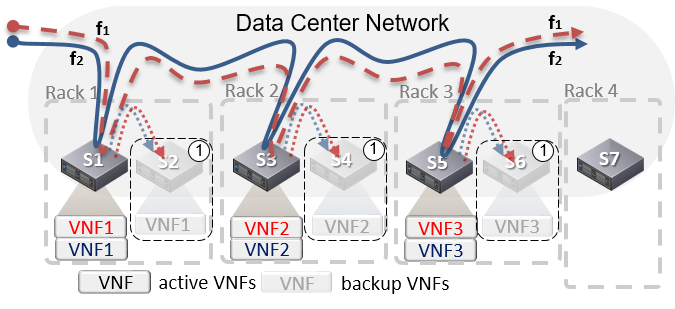}%
  \vspace{-0.2cm}
  \caption{aSbN with concentrated active and distributed backup.}
  \vspace{-0.2cm}
  \label{netC1}
\end{subfigure}
\begin{subfigure}[b]{0.9\columnwidth}
\includegraphics[width=0.9\columnwidth]{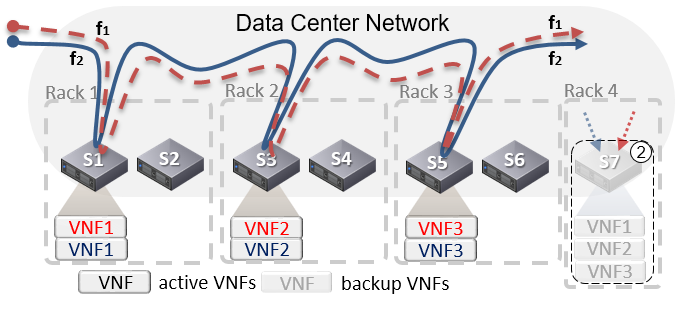}%
  \vspace{-0.2cm}
  \caption{aSbS with concentrated active and backup.}
  \vspace{-0.3cm}
  \label{netC2}
\end{subfigure}
  \caption{cVNF:Concentrated Placement of active VNFs}
  \vspace{-0.8cm}
\end{figure}

\subsubsection{cVNF: Concentrated Placement of active VNFs}
\par Fig. \ref{netC0} shows the DCN, where active VNFs of the same type are placed in the same server, e.g., VNFs of type 1 are placed in server S1, VNFs 2 - in S3, etc.; they however run on different VMs.  Thus, the parallel sub-flows share servers (S1, S3 and S5) for processing by appropriate VNFs, whereby all sub-flows experience the same end-to-end delay, as long as all VNFs are available, i.e., no OOP.

\par\textbf{aSbN}: \emph{Concentrated Active and Distributed Backup VNFs Placement:}
 In this method, as illustrated in Fig. \ref{netC1} with \circled{\small1}, the backup VNFs from backup sub-SFC are distributed over servers S2, S4 and S6.  Here, the backup server S2 protects active VNFs 1 from server S1 as well as S1 itself. When one VNFs on S1 fail, the traffic can be split between S1 and S2. In case of server, e.g., S1, failure, the service of parallel flows will be interrupted. However, if S2 can provide two backup VNFs 1, the whole traffic can be forwarded to and processed by backup server, S2, without any irruptions. When all backup VNFs on active and backup servers, e.g. on S1 and S2, or both servers fail, the service is interrupted requiring VNF migration.

\par\textbf{aSbS}: \emph{Concentrated Active and Concentrated Backup VNFs Placement:}
With this method, all active VNFs are protected by one backup server as shown in Fig. \ref{netC2} with \circled{\small2}. Here, S7 within Rack 4 is a backup server, which can generally provide one or multiple backup sub-SFCs, while recovering failures of any active VNF and even failure of active servers, e.g., S1, S3 or S5. The server and VNF protection is possible as long S7 does not fail and provides enough available backup VNFs, while backup VNFs replace only failed active VNFs. Thus, the sub-flow needs to be redirected toward backup server and then back to the active server with the next VNF from sub-SFC. That delays sub-flows and can lead to OOP. For example, when one VNF2 on server S3 fails, one sub-flow needs to be redirected to S7, whereby a new route includes S1-S3-S7-S5.

\subsubsection{dVNF: Distributed Placement of active VNFs}
\par
\begin{figure}[!t]
\centering
\begin{subfigure}[b]{0.9\columnwidth}
\includegraphics[width=0.9\columnwidth]{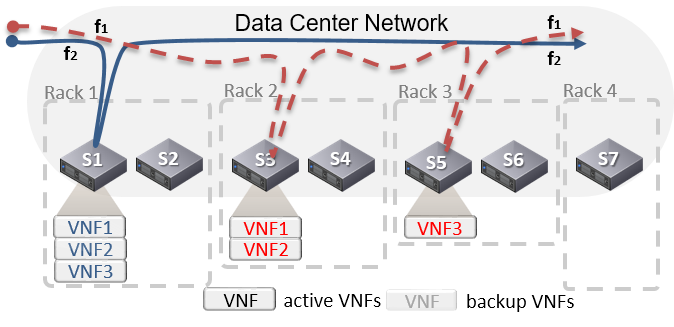}%
  \vspace{-0.2cm}
  \caption{dVNF without backup.}
  \vspace{-0.2cm}
  \label{netD0}
\end{subfigure}
\begin{subfigure}[b]{0.9\columnwidth}
\includegraphics[width=0.9\columnwidth]{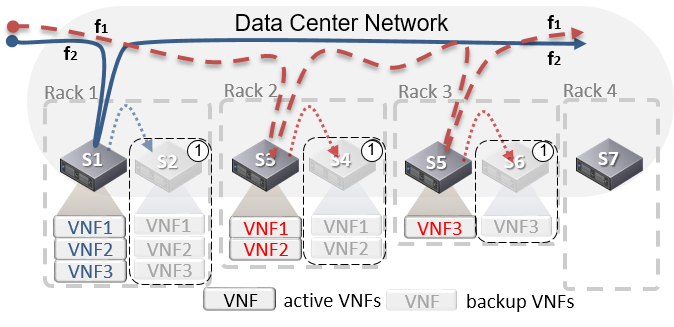}%
  \vspace{-0.2cm}
  \caption{aNbN with distributed active and backup.}
  \vspace{-0.2cm}
  \label{netD1}
\end{subfigure}
\begin{subfigure}[b]{0.9\columnwidth}
\includegraphics[width=0.9\columnwidth]{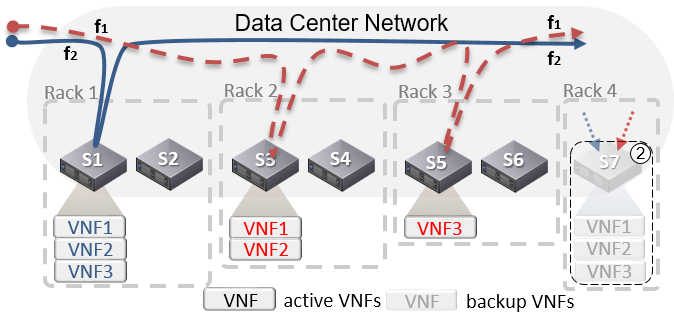}%
  \vspace{-0.2cm}
  \caption{aNbS with distributed active and concentrated backup.}
  \vspace{-0.3cm}
  \label{netD2}
\end{subfigure}
  \caption{dVNF:Distributed Placement of active VNFs}
  \vspace{-0.8cm}
\end{figure}

 Fig. \ref{netD0} shows the DCN, where VNFs of the same type are placed in different servers, e.g., VNFs of type 1 are placed in servers S1 and S3, VNFs3 in servers S1 and S5, etc.  For instance, the sub-flow $f_2$ needs to be routed only to server S1, where it passes a whole sub-SFC (VNF1-VNF2-VNF3) and finally leaves DCN. The sub-flow $f_1$ needs to pass S3 and S5 to complete the service. Generally, the distribution of VNFs from the sub-SFC can be different for different sub-flows as exemplary presented in Fig. \ref{netD0}. However, to avoid OOP, the VNFs of different parallel sub-SFC needs to be distributed in the same fashion. Thus, either the whole sub-SFC of $f_1$ needs to be placed in one server or VNF3 from sub-SFC of $f_2$ has to be placed in separate server and not in S1.

\par\textbf{aNbN}: \emph{Distributed Active and Distributed Backup VNFs Placement:}
Here, the backup sub-SFCs are distributed over multiple servers as shown by \circled{\small1} in Fig. \ref{netD1}. For example, the backup server S4 protects VNF1 and VNF2 from S3 as well as S3 itself. In contrast, S2 protects entire active sub-SFC, VNF1, VNF2 and VNF3, containing entire backup sub-SFC. That is a spatial case of dVNF in case an active sub-SFC is allocated within one server. When few VNFs on active servers fail, the traffic can be split between active and backup servers, e.g., S1 and S2. In case of server, e.g., S1, failure, the whole traffic is forwarded to and processed by backup server, S2. 
When all backup VNFs on active and backup servers, e.g. on S1 and S2, or both servers fail, the sub-flow $f_2$ can be redirected to server S4 and, then, S6 to complete the server while resulting in over-provisioning and OOP. Generally, in contrast to Fig. \ref{netD1} \circled{\small1}, not all active servers have to be replicated, e.g., the backup servers S4 and S6 could be enough for protection of both active sub-SFCs resulting however in OOP.

\par\textbf{aNbS}: \emph{Distributed Active and Concentrated Backup VNFs Placement:}
In this method, all active VNFs and all active servers can be protected by only one separate backup server. The backup server can provide one or multiple backup sub-SFCs.  As shown in Fig. \ref{netD2} by \circled{\small2}, S7 within Rack 4 is a backup server with one backup sub-SFC and can recover failures of any active VNF and even failure of active servers. For instance, when S3 and S5 fail, the service is not interrupted as long as S7 does not fail and provides entire backup sub-SFC. In case of VNF failures, only individual backup VNFs are used. Then, the sub-flows need to be redirected toward backup server and then back to the active server resulting in delays and OOP. For example, when one VNF2 on server S3 fails, $f_1$ needs to be redirected to S7, whereby a new route includes S3-S7-S5. 

\section{Analysis of reliability strategies}


The analytical models for availability and reliability for complex systems are interchangeable and only defined through availability, or reliability of the individual components \cite{REL-ETSI}. Thus, we refer to service availability (reliability) as \emph{service success}, which we define as a probability that all $n$ sub-flows of certain large flow can successfully traverse $n$ sub-SFCs. 

\par Let us denote the reliability values of active and backup servers and VM/VNFs $\varphi$, $\varphi_r$ and $\upsilon$, $\upsilon_r$ and assume that these values are the same for all servers and VNFs, respectively; and the failures of different components occur independently. Based on \cite{Lee:55}, when the same backup servers protects multiple active servers and VNFs, the availability/reliability of any backup server $\varphi_r$ may be considered only once resulting in $\varphi_r\cdot \varphi_r=\varphi_r$. That rule needs to be consider to derive equations for service success as follow 
$(a+\varphi_r b)^K\!\!=\!\!\sum_{i=0}^{K}\binom{K}{i}a^{K-i}(b\varphi_r)^i\equiv a^K\!\!+\varphi_r\sum_{i=1}^{K}\binom{K}{i}a^{K-i}b^i=a^K+\varphi_r[(a+ b)^K-a^K]$, where $a$ and $b$ are numerical values.
Additionally, we assume 
$\sum_{i=k}^{K} a_i=0\text{, if } k>K$.
\par
\begin{figure}[!t]
\centering
\includegraphics[width=1\columnwidth]{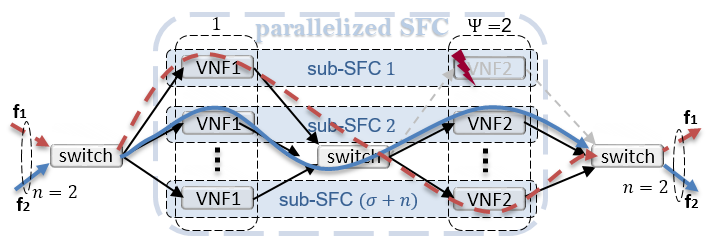}%
  \vspace{-0.2cm}
  \caption{Strategy to handle a VNF failure.}\label{vSbtVNFs}
  \vspace{-0.8cm}
\end{figure}

 Our reliability analysis is generalized by assumption that any backup VNF can replace any active VNF of the same type. Fig. \ref{vSbtVNFs} illustrates this VNF deployment strategy, which we extended to flow and SFC parallelism: DCN provides parallelized SFC, which consists of $n$, e.g., $n=2$ active sub-SFCs for processing of $n$ parallel sub-flows and additionally $\sigma$ parallel backup sub-SFCs to protect $n$ sub-SFCs. Each sub-SFC consists of $\Psi$ different VNFs, whereby each sub-flow needs to pass all $\Psi$ VNFs to complete the service. Since all $n$ sub-flows needs to be processed in the same manner with the same processing result, all $n+\sigma$ parallel VNFs of any type are synchronized and, thus, equivalent and interchangeable. As presented in Fig. \ref{vSbtVNFs}, the sub-flow $f_1$ is first sent to the VNF1 of sub-SFC 1, however, the second switch redirects it to VNF2 of sub-SFC $(\sigma+n)$ due to the failure of VNF2 of sub-SFC 1. For that scenario, where only VNFs can fail, the service success is a probability that at least $n$ over all $n+\sigma$ sub-SFCs do not fail and can serve $n$ sub-flows, while at most $\sigma$ VNFs of any type can fail without service interruption. That service success probability can be generalized as
$R(n)=[\sum_{f=0}^{\sigma}\binom{\sigma+n}{f}(1-\upsilon)^f\upsilon^{\sigma+n-f}]^{\Psi}$.
Using this general expression, we derive the service success $R(n)$ for each defined VNFs placements, where we consider both VNF and server failures. The notations are summarized in Table \ref{t1}.

\subsection{Analysis on Service Success Probability}

\subsubsection{cVNF:Concentrated Placement of active VNFs}

Here, each active server contains the same type of VNFs, whereby $N=\Psi$ servers are required to implement $n$ sub-SFCs. Thus, each out of $n\geq1$ sub-flows needs to pass all $\Psi$ servers. When there is no failure protection of parallelized SFC, the service success is defined as
\begin{equation}
R_{0,\text{cv}}(n)=[\varphi\upsilon^{n}]^{\Psi},
\end{equation}
whereby all $\Psi$ servers and all $n\Psi$ VNFs of $n$ sub-SFCs have to be available.

\textbf{aSbN:}
Each out of $\Psi$ active servers and its VNFs can be protected by one backup server resulting in $m=\Psi$ backup servers. Each backup server contains $\sigma\geq1$ backup VNFs of certain type resulting in $\sigma\geq1$ backup sub-SFCs over DCN. When active server is available, at most $i=min\{n,\sigma\}$ active VNFs and $\sigma-i$ backup VNFs can fail on active and backup servers without service interruption, respectively. When active server fails and backup server is available and contains $\sigma\geq n$ backup VNFs, the service success can be still provided. In this case, at most $\sigma-n$ VNFs on any backup server can fail without service interruption, i.e., the service success probability is 
\begin{small}
\begin{equation}\label{2a}
\begin{split}
&R^{\small1}_\text{cv}(n)=\bigg[\varphi\upsilon^{n}+\varphi \varphi_r \sum_{i=1}^{min\{n,\sigma\}}\binom{n}{i}(1-\upsilon)^i \upsilon^{n-i}  \sum_{j=0}^{\sigma-i}\binom{\sigma}{j}\\ &\cdot\upsilon_r^{\sigma-j}(1-\upsilon_r)^j+(1-\varphi)\varphi_r\sum_{j=0}^{\sigma-n}\binom{\sigma}{j}(1-\upsilon_r)^j\cdot\upsilon_r^{\sigma-j}\bigg]^{\Psi}
\end{split}
\end{equation}
\end{small}
\vspace{-0.4cm}

\textbf{aSbS:}  
One backup server can contain $\sigma\geq1$ backup sub-SFCs and is utilized, when active server or some active VNFs fail. When all $N=\Psi$ active servers are available, at most $i\in[0,min\{n,\sigma\}]$ active VNFs of each type and $j=[0,\sigma-i]$ backup VNFs of the same type can fail without service interruption. Generally, $f\in[0,\Psi]$ active servers can fail without service interruption, when backup server provides $\sigma\geq n$ backup sub-SFCs. Then, any server failure results in failure of $n$ active VNFs of the same type, which are replaced by $n$ backup VNFs, whereby $l\in[0,\sigma-n]$ backup VNFs of the same type on backup server can also fail. The active VNFs on the remaining $\Psi-f$ active servers can fail as well, while at most $min\{n,\sigma\}$ active VNFs of each type can fail and be replaced by backup VNFs. Thus, the service success is
\begin{small}
\begin{equation}
\begin{split}
&R^{\small2}_\text{cv}(n)=\varphi^{\Psi}\upsilon^{n\Psi}   +\varphi^{\Psi}\varphi_r\bigg(\bigg[\sum_{i=0}^{min\{n,\sigma\}}\binom{n}{i}(1-\upsilon)^i\upsilon^{n-i}
\\ & \sum_{j=0}^{\sigma-i}\binom{\sigma}{j}(1-\upsilon_r)^j\upsilon_r^{\sigma-j} \bigg]^{\Psi}-\upsilon^{n\Psi}\bigg)+ \varphi_r\cdot
\\&\sum_{f=1}^{\Psi}\binom{\Psi}{f} \varphi^{\Psi-f}(1-\varphi)^f  \bigg[\sum_{l=0}^{\sigma-n}\binom{\sigma}{l}\upsilon_r^{\sigma-l}(1-\upsilon_r)^l\bigg]^f 
\\&\cdot\bigg[\sum_{i=0}^{min\{n,\sigma\}}
\!\!\!\!\binom{n}{i}
(1-\upsilon)^i\upsilon^{n-i}
\sum_{j=0}^{\sigma-i}\binom{\sigma}{j}(1-\upsilon_r)^j\upsilon_r^{\sigma-j}\bigg]^{\Psi-f}
\end{split}
\end{equation}
\end{small}
\vspace{-0.4cm}
\subsubsection{dVNF: Distributed Placement of active VNFs }
Without any failure protection, $N\geq1$ active servers allocate one activ sub-SFC to serve one out of $n$ sub-SFCs to process one out of $n$ sub-flows resulting in $nN$ active servers to allocate $n$ sub-SFCs. Any active server $k\in[1,N]$ of a certain sub-flow contains $\psi_k\leq\Psi$, $\sum_{k=1}^{N}\psi_k=\Psi$, different types of VNFs. Thus, the service success probability is determined as 
\begin{equation}
R_{0,\text{dv}}(n)=\big[\varphi^N\prod_{k=1}^{N}\upsilon^{\psi_k}\big]^n
\end{equation}
\vspace{-0.3cm}
\begin{table}[]
\centering
\caption{Notation}\label{t1}
\vspace{-0.3cm}
\resizebox{\columnwidth}{!}{
\begin{tabular}{|l|l|}
\cline{1-2}
 $\varphi$ & availability/reliability of the active server; \\ 
 $\varphi_r$&  availability/reliability of a backup server;   \\
 $\upsilon$&  availability/reliability of VNF on active server;  \\
$\upsilon_r$&  availability/reliability of VNF on backup server;  \\
 $\Psi$ &   number of VNFs in a SFC and, thus, sub-SFC; \\
$n$ &   number of parallel sub-flows and, thus, active sub-SFCs; \\
 $N$&  number of active servers per sub-flow;  \\
 $\psi_k$&  number of VNFs from the same sub-SFC on server $k\in[1,N]$;  \\
 $\sigma$& number of backup VNFs of the same type, i.e, backup sub-SFCs;\\
 $\sigma_{_{\sum}}$& a total number of backup sub-SFCs;\\
 $m$ & number of backup servers;  \\\cline{1-2}
\end{tabular}
}
\vspace{-0.6cm}
\end{table}

\textbf{aNbN:} When $m$, $N\leq m\leq nN$, backup servers contain $\sigma\geq1$ backup VNFs of certain type, there are $\tfrac{m}{N}\sigma$ backup sub-SFCs to protect $n$ sub-flows. Generally, each failed VNF of certain type can be replaced by any out of $\tfrac{m\sigma}{N}$ backup VNFs of the same type through redirection of individual sub-flows. Let's consider all $nN$ active and $\tfrac{m}{N}$ backup servers of type $k$ over all $n$ sub-flows. We assume that at least one out of $\tfrac{m}{N}$ backup servers have to be available to recover failures of active servers or VNFs. When $l\in[0,\tfrac{m}{N}-1]$ backup servers fail and all $n$ active servers of type $k$ are available, $i\in[0,min\{n,(\tfrac{m}{N}-l)\sigma\}]$ active VNFs of certain type over all $n$ active servers can fail without service interruption. In this case, available backup servers have to provide enough backup VNFs of the same type, while $j=\in[(\tfrac{m}{N}-l)\sigma-i]$ backup VNFs can fail without impact on service success. 
When $f\in[1,min\{n,(\tfrac{m}{N}-l)\sigma\}]$ active servers fails, it is allowed that 
 $i\in[0,min\{n-f,(\tfrac{m}{N}-l)\sigma-f\}]$ additional active VNFs fail among $n-f$ available active servers. The $i$ failed active VNFs are recovered by backup VNFs of the same type from $\tfrac{m}{N}-l$ backup servers, whereby $(\tfrac{m}{N}-l)\sigma-f-i$ backup VNFs of the same type can fail without service interruption.
By considering that each active server provides $\psi_k$ different VNFs and that each sub-flow has to pass $N$ servers, the service success is defined as
\begin{small}
\begin{equation}\label{multibackup}
\begin{split}
&R^{\small1}_\text{dv}(n)=\prod_{k=1}^{N}\bigg((\varphi\upsilon^{\psi_k})^n+\sum_{l=0}^{\tfrac{m}{N}-1}\binom{\tfrac{m}{N}}{l}\varphi_r^{\tfrac{m}{N}-l}(1-\varphi_r)^l
\\&\cdot\bigg\{\varphi^n \bigg(\bigg[\sum_{i=0}^{min\{n,(\tfrac{m}{N}-l)\sigma\}}\binom{n}{i}(1-\upsilon)^i\cdot
 \upsilon^{n-i} \sum_{j=0}^{(\tfrac{m}{N}-l)\sigma-i}
 \\&\binom{(\tfrac{m}{N}-l)\sigma}{j}(1-\upsilon_r)^j \upsilon_r^{(\tfrac{m}{N}-l)\sigma-j}\bigg]^{\psi_k}-(\upsilon^{\psi_k})^n\bigg)
\\&+\sum_{f=1}^{min\{n,(\tfrac{m}{N}-l)\sigma\}}
\binom{n}{f}\varphi^{n-f}(1-\varphi)^f \bigg[\sum_{i=0}^{min\{n-f,(\tfrac{m}{N}-l)\sigma\}-f}\\& \binom{{n-f}}{i}
(1-\upsilon)^i\upsilon^{{n-f}-i}\!\!\!\sum_{j=0}^{(\tfrac{m}{N}-l)\sigma-f-i}\binom{(\tfrac{m}{N}-l)\sigma}{j}(1-\upsilon_r)^j
\\&\upsilon_r^{(\tfrac{m}{N}-l)\sigma-j}\bigg]^{\psi_k}\bigg\}\bigg),
\end{split}
\end{equation}
\end{small}

\textbf{aNbS:}  
When one backup server contains $\sigma\geq1$ backup sub-SFCs, $f\in[0,min\{n,\sigma\}]$ active servers and $i\in[0,min\{n-f,\sigma-f\}]$ active VNFs of each type can fail without service interruption. Thus, the backup server has to be available to provide at least $f+i$ backup VNFs of certain type, while $j\in[0,\sigma-f-i]$ backup VNFs can still fail. The service success probability is then determined as
\begin{small}
\begin{equation}
\begin{split}
&R^{\small2}_\text{dv}(n)=\prod_{k=1}^{N}\bigg((\varphi\upsilon^{\psi_k})^n+\varphi^n \varphi_r\bigg(\bigg[\sum_{i=0}^{min\{n,\sigma\}}\binom{n}{i}(1-\upsilon)^i\cdot
\\&\cdot \upsilon^{n-i} \sum_{j=0}^{\sigma-i}\binom{\sigma}{j}(1-\upsilon_r)^j \upsilon_r^{\sigma-j}\bigg]^{\psi_k}-(\upsilon^{\psi_k})^n\bigg)
+\sum_{f=1}^{min\{n,\sigma\}}
\\&\binom{n}{f}\varphi^{n-f}(1-\varphi)^f \varphi_r \bigg[\sum_{i=0}^{min\{n-f,\sigma-f\}}\binom{{n-f}}{i}\cdot
\\& \cdot(1-\upsilon)^i\upsilon^{{n-f}-i}\sum_{j=0}^{\sigma-f-i}\binom{\sigma}{j}(1-\upsilon_r)^j\upsilon_r^{\sigma-j}\bigg]^{\psi_k}\bigg)
\end{split}
\end{equation}
\end{small}

\subsection{Overhead Analysis}
We analyze overhead by considering utilization of reserved VNFs. We define \emph{resource utilization} $\Omega$ as ratio between the number of  VNFs required for implementation of $n$ sub-SFCs and the amount of all reserved backup and active VNFs. Generally, to maintain the service of all $n$ parallel sub-flows, at least $n$ sub-SFCs are required. Since each sub-SFC consists of $\Psi$ VNFs, the total number of VNFs required is $n\Psi$ VNFs.


\textbf{aSbN}
When $\Psi$ active servers are protected by $\Psi$ backup servers, while each of them contains $\sigma$ VNFs, there are in total $\sigma\Psi$ backup VNFs ($\sigma$ backup sub-SFCs) to protect $n$ sub-flows and resource utilization is 
\begin{equation}
\Omega^{\small1}_\text{cv}=\tfrac{n\Psi}{\Psi(n+\sigma)}=\tfrac{n}{n+\sigma}
\end{equation}
\vspace{-0.6cm}

\textbf{aSbS:}  If a single backup server reserves $\sigma$ sub-SFCs with $\Psi$ VNFs each, the total amount of reserved VNFs is $\Psi(n+\sigma)$ and the resource utilization is
\begin{equation}\label{OmegaCV2b}
\Omega^{\small 2}_\text{cv}=\tfrac{n\Psi}{\Psi(n+\sigma)}=\tfrac{n}{n+\sigma}
\end{equation}
\vspace{-0.6cm}

\begin{figure}[!t]
\centering
\includegraphics[width=0.85\columnwidth]{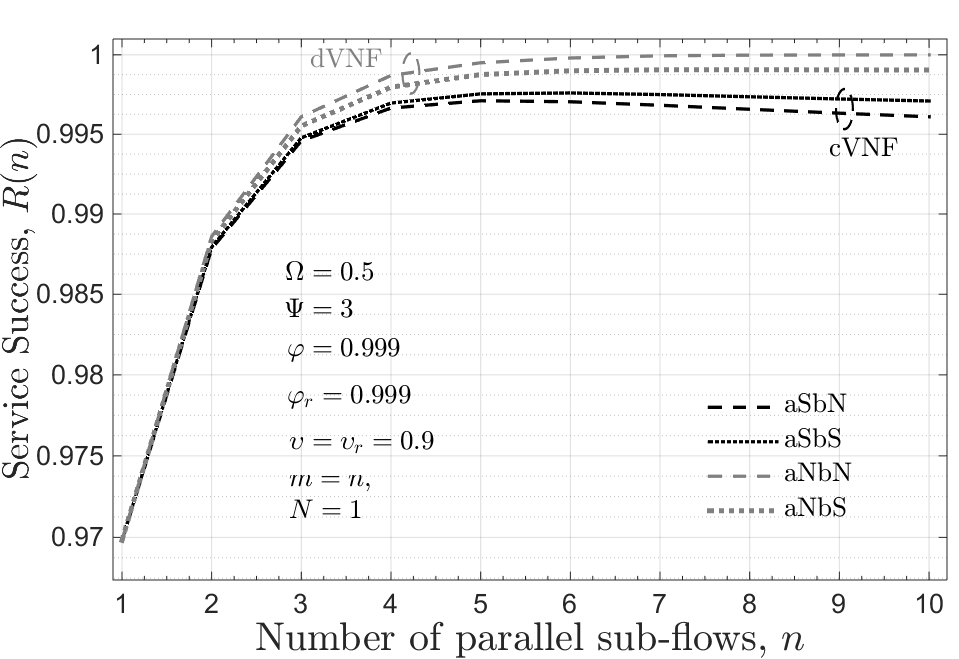}%
  \vspace{-0.2cm}
  \caption{\small Service success vs. number of parallel sub-flows $n$}
  \vspace{-0.5cm}
  \label{f1}
\end{figure}

\textbf{aNbN:}When $nN$ active servers are protected by $m\geq N$ backup servers, which contain $\sigma$ backup VNFs of certain type, i.e., $\sigma$ sub-SFCs with $\Psi$ VNFs, $\tfrac{m}{N} \sigma$ backup sub-SFCs protect $n$ active sub-SFCs resulting in $\tfrac{m\sigma\Psi}{N}$ backup VNFs. Then, the resource utilization is 
\begin{equation}
\Omega^{\small1}_\text{dv}=\tfrac{n\Psi N}{n\Psi N+m\sigma\Psi}=\tfrac{nN}{nN+m\sigma}
\end{equation}\vspace{-0.6cm}

\textbf{aNbS:}
When single backup server reserves $\sigma$ sub-SFCs with $\Psi$ VNFs each, the resource utilization is defined by Eq. \eqref{OmegaCV2b}, i.e., $\Omega^{\small2}_\text{dv}=\Omega^{\small2}_\text{cv}$.

\section{Performance Evaluation}
Now, we evaluate the service success probability $R(n)$ and resource utilization $\Omega$ of the proposed methods for VNF placement and protection, whereby we verify analytical results by Monte-Carlo simulations. Since the simulation results overlapped with analytical results and were obtained with 95\% confidence interval, we demonstrate only analytical results. Since we generally assume, that 
VNF failure is more likely than server failure, we set availability/reliability of any server and any VNF as $\varphi=\varphi_r=0.999$ and $\upsilon=\upsilon_r=0.9$, respectively. As expected, VNF protection method dVNF shows the best results for $N=1$, i.e., we set $N=1$ for evaluation below.
 
\par Fig. \ref{f1} shows the service success $R(n)$ as a function of a number of parallel sub-flows $n$ and predefined resource utilization $\Omega$. The amount of backup VNFs, $\sigma$, and servers $m$ for each protection method was set so that the resulting resource utilization $\Omega$ reaches $0.5$, i.e., 1:1 protection. The service success increases up to $100\%$ and $99.9\%$ with increasing number of parallel sub-flows $n$ in case of dVNF: aNbN and aNbS, respectively. In case of cVNF: aSbN and aSbS, the maximal service success of $99.71\%$ and $99.76\%$ was reached with $n=5$ and $n=6$ sub-flows/sub-SFCs, respectively. 
\begin{figure}[!t]
\centering
\includegraphics[width=0.85\columnwidth]{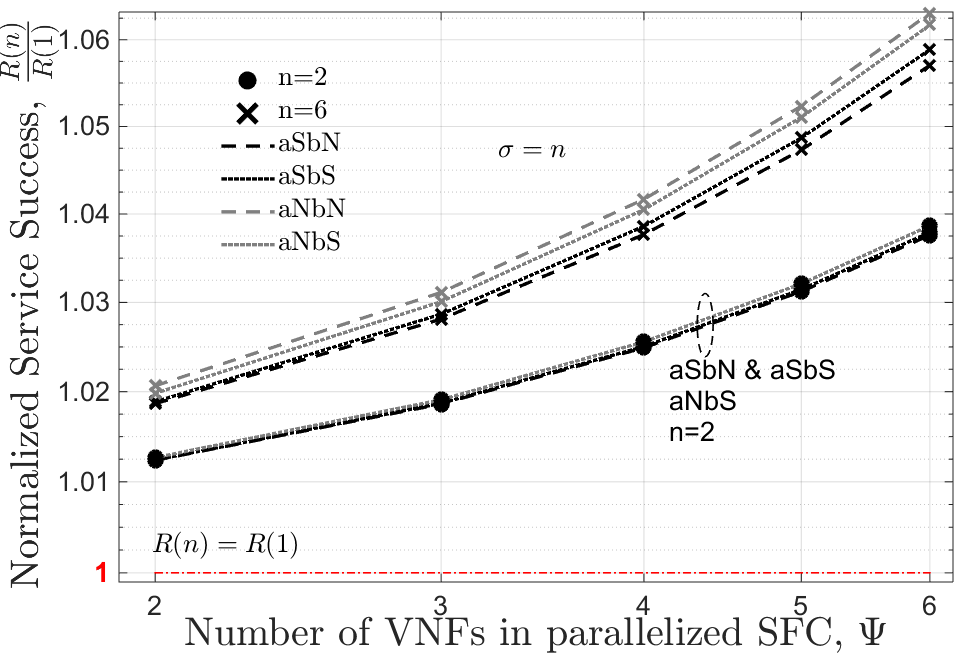}%
  \vspace{-0.3cm}
  \caption{\small Normalized service success vs. amount of VNFs in SFC.}
  \vspace{-0.6cm}
  \label{f3}
\end{figure}

\par Fig. \ref{f3} demonstrates the normalized service success as a function of number of VNFs in SFC $\Psi$ and the number of parallel sub-flows $n$. We set the number of backup sub-SFCs as $\sigma=n$, that allows to maintain the service of all $n$ sub-flows as long as at least one server is available. The service success probability for $n>1$ is normalized by the service success probability of serial flow, i.e., $n=1$. We studied the service success for $n=2$ and $n=6$, but the results for $n=2$ are only shown for aSbN and aSbS and aNbS due to significant deviation from results for $n=6$. Generally, all protection methods with $n>1$ outperformed service success of serial flow and showed increasing service success with increasing number of VNFs in SFC and amount of sub-SFCs $n$.

\begin{table}[t]
\centering
\caption{\small Required amount of backup VNFs ($\sigma$) and resource utilization ($\Omega$) for the service success of $R(n)\geq 0.999$}\vspace{-0.2cm}
\label{tt}
\resizebox{0.93\columnwidth}{!}{%
\begin{tabular}{|c|c|c|c|c|c|c|c|c|c|c|}
\hline
\multirow{3}{*}{\begin{tabular}[c]{@{}c@{}}VNF \\ Protection\end{tabular}} & \multicolumn{5}{c|}{cVNF: \textbf{aS-}}         & \multicolumn{5}{c|}{dVNF: \textbf{aN-}}              \\ \cline{2-11} 
& \multicolumn{2}{c|}{$n=1$} & \multicolumn{3}{c|}{$n>1$} & \multicolumn{2}{c|}{$n=1$} & \multicolumn{3}{c|}{$n>1$}     \\ \cline{2-11} 
& $\sigma_{_{\sum}}$     & $\Omega$    & $n$  & $\sigma_{_{\sum}}$ & $\Omega$ & $\sigma_{_{\sum}}$     & $\Omega$    & $n$      & $\sigma_{_{\sum}}$ & $\Omega$ \\ \hline
 $\textbf{-bN}$                                                                  & 3            & 0.25        & 3    & 5        & 0.375    & 3            & 0.25        & $\geq15$ & $\geq8$        & $\geq0.652$      \\ \hline
 $\textbf{-bS}$                                                                  & 3            & 0.25        & 6    & 8        & 0.429    & 3            & 0.25        & $\geq 9$        & $\geq 8$        & $\geq 0.529$    \\ \hline
\end{tabular}%
}\vspace{-0.4cm}
\end{table}
Table \ref{tt} demonstrates the total number of backup sub-SFCs $\sigma_{_{\sum}}$, resource utilization ($\Omega$) for serial $n=1$ and parallel $n>1$ SFCs and for each proposed VNF placement of active and backup VNFs, when the sub-SFCs consist of $\Psi=3$ VNFs and the probability for service success has to be at least $R(n)\geq 0.999$. In case of aNbD, the number of backup servers was set to $m=2$, $N=1$ for $n>1$. The parallel VNF chaining outperforms serial SFC regarding resource utilization $\Omega$, which can be increased by increasing the number of parallel sub-flows $n$. Especially, the resource utilization of dVNF: aNbN and aNbS can be further increased, e.g., $\Omega^{\small2}_\text{dv}\geq0.529$ and $\Omega^{\small1}_\text{dv}\geq0.652$ by increasing $n$ and $\sigma$. Generally, the parallelism applied to dVNF can at least double the resource utilization $\Omega$ for both protection methods, whereby distributed placement of backup servers, aNbN, shows the best performance, e.g., $\sigma_{_{\sum}}=8$ backup VNFs can protect $n\leq15$ parallel sub-flows. However, we would like to note that only this aNbN could reach the service success probability up to $R(n)\geq0.99999$. 

\begin{figure}[!t]
\centering
\includegraphics[width=0.85\columnwidth]{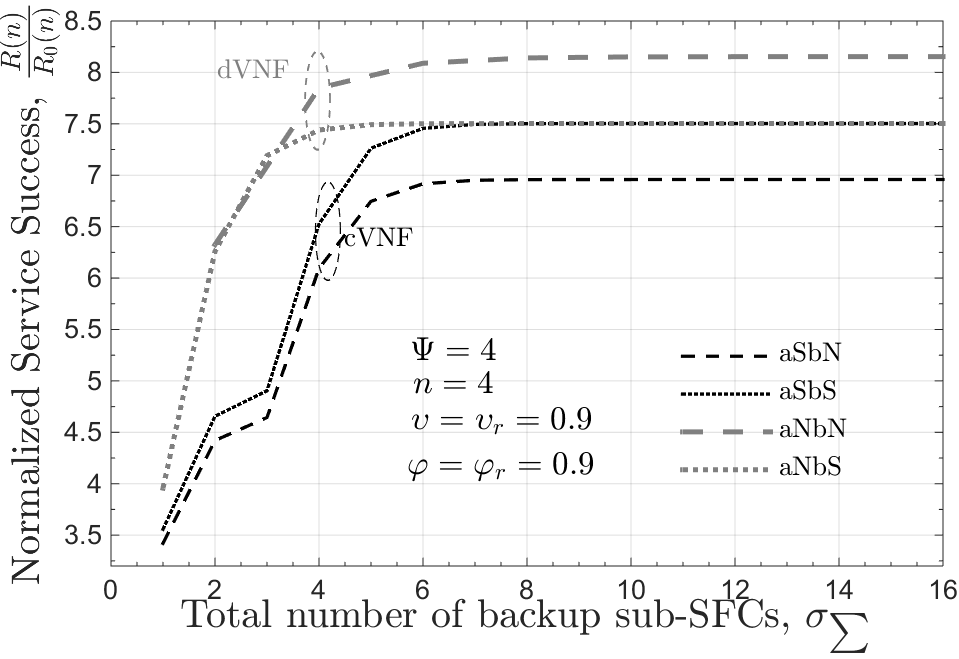}%
  \vspace{-0.3cm}
  \caption{\small Service success vs. a total amount of backup VNFs.}
  \vspace{-0.8cm}
  \label{f5}
\end{figure}

Fig. \ref{f5} shows the service success related to each VNF placement and protection method normalized by service success without any protection $R_0(n)$. We set $\varphi=\varphi_r=\upsilon=\upsilon_r=0.9$, the number of VNFs in SFC as $\Psi=4$ and the number of sub-flows as $n=4$ to reach the same service success for cVNF and dVNF, $R_0(n)=R_{0,\text{cv}}(n)=R_{0,\text{dv}}(n)$. For that settings, we measured during simulation \emph{the maximal number of inter-rack hops} of sub-flows in case of service success with a result of 5, 9,13 and 9 inter-rack hops for aSbN, aSbS, aNbN and aNbS, respectively. The aSbN shows the minimal number of inter-rack hops, i.e., 5, while the worst performance regarding service success improving it only by factor 7 with a large number of backup sub-SFCs, $\sigma_{_{\sum}}>6$. In contrast, aNbN improves the service success by factor 8 already with $\sigma_{_{\sum}}\geq4$ resulting in the maximal number of inter-rack hops.


\section{Conclusion}
We studied end-to-end service reliability in DCNs with flow and SFC parallelism, where any large flow is split into multiple parallel smaller sub-flows and any SFC is replicated into multiple sub-SFCs. We defined two placement methods for active VNFs, where VNFs of the same type are distributed over DCN, dVNF, or concentrated in the same servers, cVNF, and compared two different placement strategies for backup VNFs applied to dVNF and cVNF, whereby all backup sub-SFCs can be concentrated in one backup server or distributed over multiple backup servers.  
Based on the probability theory, we analytically derived for each studied VNF placement method the service reliability in case of server and VNF failures as a function of flow and SFC parallelism and placement of parallel active and backup sub-SFCs within DCN.  
The results showed that the parallelism in DCN significantly increases service reliability up to $0.99999$, while requires much less, at least $33\%$ less, backup VNFs as compared to service of serial traffic flow. Especially dVNF with $8$ backup sub-SFCs distributed over $2$ backup servers can protect up to $15$ sub-SFCs with reliability $0.999$ and efficiency of resource utilization $65.2\%$.

\bibliographystyle{IEEEtran}
\bibliography{bibL}
\end{document}